\documentclass{article}
\usepackage{authblk}
\usepackage[numbers]{natbib}
\usepackage{rotating}
\usepackage{graphicx}
\usepackage{subfigure}
\usepackage[utf8]{inputenc}
\usepackage{amssymb}
\usepackage[english]{babel}
\usepackage{multicol}
\usepackage{color}
\usepackage{xspace}

\begin{document}

\title{Ligament break-up simulation through pseudo-potential Lattice Boltzmann Method}

\author[1]{Daniele Chiappini}
\author[2,4]{Xiao Xue}
\author[3]{Giacomo Falcucci}
\author[2]{Mauro Sbragaglia}
\affil[1]{University of Rome 'Niccolò Cusano', Department of Industrial Engineering, Via don Carlo Gnocchi 3, 00166, Rome, Italy, daniele.chiappini@unicusano.it}
\affil[2]{University of Rome 'Tor Vergata', Department of physics and INFN, Via della Ricerca Scientifica 1, 00133 Rome Italy}
\affil[3]{University of Rome 'Tor Vergata', Department of Enterprise Engineering “Mario Lucertini", Via del Politecnico 1 1, 00133 Rome Italy}
\affil[4]{Department of Physics, Eindhoven University of Technology, 5600 MB Eindhoven}

\maketitle

\begin{abstract}

The Plateau-Rayleigh instability causes the fragmentation of a liquid ligament into smaller droplets. In this study a numerical study of this phenomenon based on a single relaxation time (SRT) pseudo-potential lattice Boltzmann method (LBM) is proposed. If systematically analysed, this test case allows to design appropriate parameters sets to deal with engineering applications involving the hydrodynamics of a jet. Grid convergence simulations are performed in the limit where the interface thickness is asymptotically smaller than the characteristic size of the ligament. These simulations show a neat asymptotic behaviour, possibly related to the convergence of LBM diffuse-interface physics to sharp interface hydrodynamics.

\end{abstract}

\section{INTRODUCTION}
Numerical methods coupled with experimental tests are gaining nowadays great importance for solving a large number of problems which involve both innovative applications (from aerodynamics to multiphase) or innovative materials, \cite{Vesselin2011,Fanelli2017a,Fanelli2017}.
In this scenario, during the recent years an increasing attention has been driven towards the lattice Boltzmann method (LBM) for solving many physical problems \cite{Succi, Qian, Higuera, Ansumali, Benzi, Aidun2010}. The major appeals pertain its simplicity and applicability in a wide range of conditions, the easy handling of complex geometries (e.g. porous media), the possibility to combine it with other methods to design hybrid approaches \cite{:Chiappini2015, DiIlio2016b}, and even possible extensions to more complex physics phenomena (e.g. non-Newtonian fluids, \cite{DiIlio2016a}). In this ballpark, one of the most common (modern) application, where to exploit all the advantages related to such a method, is multiphase fluid dynamics. LBM, in fact, allows to locally solve the equation of state with apparent benefits for understanding physical phenomena, \cite{:ShanChen, Sbragaglia2006, Lee2006a, Falcucci, Sbragaglia2006a, Kupershtokh2009, Falcucci2010, 2010-01-1130,Colosqui2012}. Many interesting engineering applications have been already analysed with different multiphase approaches, \cite{Chiappini, Bella2009a, Chiappini2009, Falcucci2006, Falcucci2009, Falcucci, Falcucci2012}. The aim of this work is to deeply analyse the Plateau-Rayleigh (PR) instability by means of a single-belt pseudo-potential Lattice Boltzmann method. LBM is by definition a diffuse interface method: at the interface between two different phases, the fluid properties (i.e. density, velocity, pressure) change smoothly over a region with assigned thickness. With regard to the PR instability, this sets a compelling case for conducting a series of numerical simulations in order to understand if an asymptotic behaviour may be recognized when the interface thickness is smaller with respect to the characteristic geometrical length-scale of the physical problem at hand.

\section{NUMERICAL MODEL}
The Lattice Boltzmann Method represents a relatively new computational approach which may be alternative to standard Navier-Stokes based solvers. It starts from a microscopic kinetic approach in order to reconstruct the macroscopic fluid flow solutions by mathematically describing movements and interactions of the particles which constitute the flow. The LBM \cite{Succi, Benzi, Aidun2010} starts from the Boltzmann Equation and, after a discrete decomposition, it writes the discrete form of the Boltzmann equation itself for a fixed set of speeds. The discrete Boltzmann equation reads as follows:
\begin{equation}
	\frac{\partial f_\alpha\left( \textbf{x},t \right)}{\partial t}+\textbf{c}_{\alpha} \nabla f_\alpha\left(\textbf{x},t\right) = -\frac{1}{\tau}\left[f_\alpha\left(\textbf{x},t\right) - f_\alpha^{eq}\left(\textbf{x},t\right)\right]
	\label{eq:LBE}
\end{equation}
where $f\left(\textbf{x},t\right)$ is the particle distribution function which is generally solved along the allowed velocities directions, \cite{Aidun2010}. In this paper a $D3Q19$ stencil is used for flow discretization coupled with a uniform Cartesian lattice. The single-relaxation time $\tau$ to the local equilibrium $f_{eq}$ , which is a function of the macroscopic flow variables according to the following equation \ref{eq:local_eq}, is used:
\begin{equation}	
	\label{eq:local_eq}
	f_\alpha^{eq}\left(\textbf{x},t\right) =  w_\alpha\rho\left(\textbf{x},t\right) \bigg[\frac{\textbf{c}_\alpha\cdot\textbf{u}\left(\textbf{x},t\right)}{c_s^2}+\frac{\left[\textbf{c}_\alpha\cdot\textbf{u}\left(\textbf{x},t\right)\right]^2}{2c_s^4}  - \frac{\left[\textbf{u}\left(\textbf{x},t\right)\cdot\textbf{u}\left(\textbf{x},t\right)\right]}{2c_s^2}\bigg]
\end{equation}
where $w_{\alpha}$ represents a set of weights normalized to unity; $\rho\left(\textbf{x},t\right)$ and $\textbf{u}\left(\textbf{x},t\right)$ are respectively global density and speed and $c_s$ is the lattice speed of sound.\\
Equation \ref{eq:LBE} may be explicitly solved as follows:
\begin{equation}
	\label{eq:DBE}
		\Delta_\alpha f_\alpha\left(\textbf{x},t\right) = f_\alpha\left(\textbf{x}+\textbf{c}_\alpha\Delta t,t+\Delta t\right)-f_\alpha\left(\textbf{x},t\right) = -\frac{\Delta t}{\tau}\left[f_\alpha\left(\textbf{x},t\right) - f_\alpha^{eq}\left(\textbf{x},t\right)\right]+ F_{\alpha}
\end{equation}
where $F_{\alpha}$ represents the external forcing - in this case the multiphase interaction described below.
Macroscopic fluid density and speed may be derived respectively through the $0^{th}$ and the $1^{st}$ population momentum, as in the following equation:
\begin{equation}
	\label{eq:momentum}
	\rho\left(\textbf{x},t\right)=\sum_{\alpha=0}^{N_{pop}-1}f_\alpha\left(\textbf{x},t\right) \;\;\;\;\;\;\;\;\;\;\;	\rho\textbf{u}\left(\textbf{x},t\right)=\sum_{\alpha=0}^{N_{pop}-1}\textbf{c}_\alpha f_\alpha\left(\textbf{x},t\right)
\end{equation}
where $N_{pop}$ denotes the number of discrete velocities (or populations).
As previously anticipated, this work is based on the pseudo-potential forcing which, in case of single-belt formulation \cite{:ShanChen}, may be written as:
\begin{equation}\label{SC_force}
  \textbf{F}\left(\textbf{x},t\right) = -{G_0}\psi \left(\textbf{x},t\right)\sum_{\alpha}^{N_{pop}}\psi \left(\textbf{x}+\textbf{c}_{\alpha}\Delta t,t\right)\textbf{c}_{\alpha}w_{\alpha}
\end{equation}
being $\psi\left(\textbf{x},t\right)$ a local functional of a density: $\psi\left(\textbf{x},t\right) = \rho_0\left[1 - \exp\left(-{\rho\left(\textbf{x},t\right)\over{\rho_0}}\right)\right]$. In this application the reference density $\rho_0$ is set to $\rho_0=1$ and $G_0$ is the basic parameter which rules the inter-particle interaction. By the assumption on $\rho_0$, the inter-particle strength $G_0$ is the only free parameter which fixes both the density ratio and the surface tension. Starting from equation \ref{SC_force}, the component of the interaction potential along each direction can be evaluated and then used to shift the macroscopic velocities before evaluating the equilibrium distribution functions:
\begin{equation}
\label{vel_shift}
	\textbf{u}^{'}\left(\textbf{x},t\right) = \textbf{u}\left(\textbf{x},t\right) + {{\textbf{F}\left(\textbf{x},t\right)\tau}\over{\rho\left(\textbf{x},t\right)}}
\end{equation}
Finally, the equation of state of the system may be written as follows:
\begin{equation}\label{EOS_SC}
	P=p_0+{{c_s^2G_0}\over{2}}\psi^2=\rho c_s^2 + {{c_s^2G_0}\over{2}}\psi^2
\end{equation}
\section{LIGAMENT SET-UP}
\begin{figure}
\centering
\includegraphics[width=6cm]{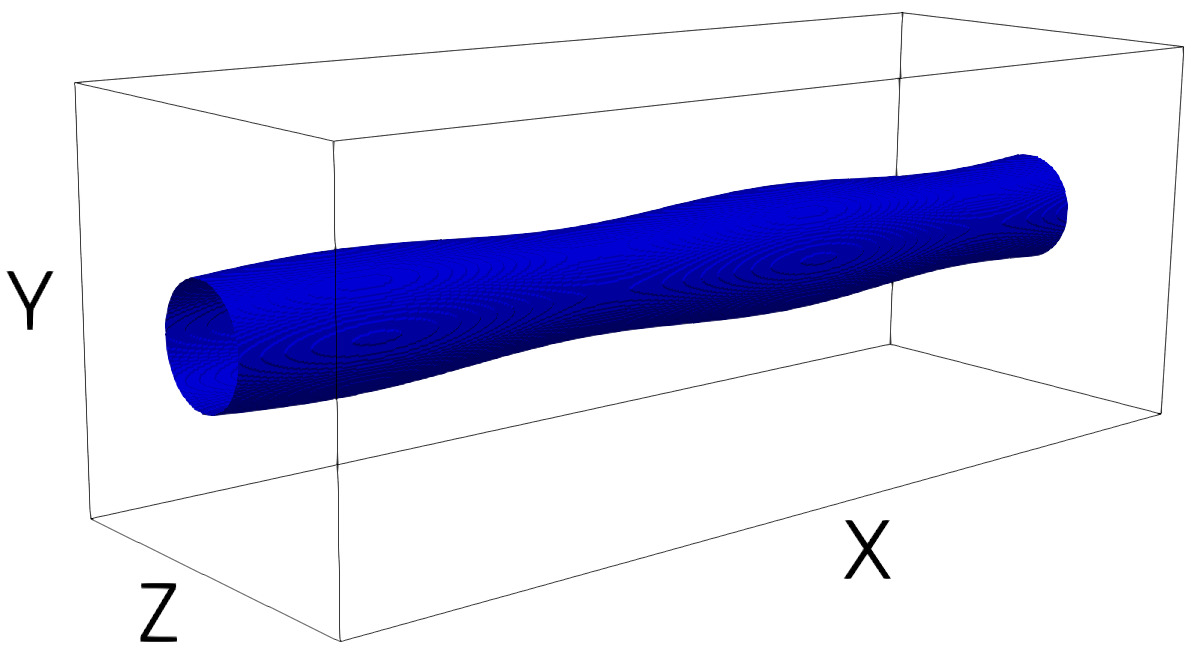}
\caption{Perturbation of ligament radius along $x$ coordinate.}
\label{fig:ligament_init}
\end{figure}

In this section the test case and the main parameters are presented. First of all it is useful to define the initial ligament configuration in order to reach its breakup. For the Plateau-Rayleigh instability a sinusoidal perturbation has used with a fixed amplitude and a constant frequency. More specifically the perturbation amplitude is a function of the ligament radius, while the period of oscillation is related to the domain length.  With referring to Fig. \ref{fig:ligament_init}, the radius $r$ function of the $x$ coordinate may be written as: $r\left(x\right) = R_0 + \frac{R_0}{10}\sin\left(\frac{4x}{L}\right)$, where $R_0$ is the unperturbed ligament radius and $L$ is the domain length.
The phenomenon is driven by some characteristic parameters, between them the Ohnesorge number $Oh$ and the capillary time $t_{cap}$ defined in the following equation:
\begin{equation}
Oh = \frac{\nu}{\rho_l\sigma\hat{r}} \;\;\;\;\;\;\;\; t_{cap} = \sqrt{\frac{\rho_l\hat{r}^3}{\sigma}}
\end{equation}
where $\rho_l$ is the liquid density, $\nu$ the kinematic viscosity, $\sigma$ the surface tension, $\hat{r}$ the average radius of the ligament along $x$ coordinate. 
Starting from this configuration and with the parameters above defined, the dimensionless time of break up as a function of initial radius will be analysed.
\section{RESULTS}
\begin{figure}
\begin{minipage}{0.55\textwidth}
\centering
\includegraphics[width=0.95\columnwidth]{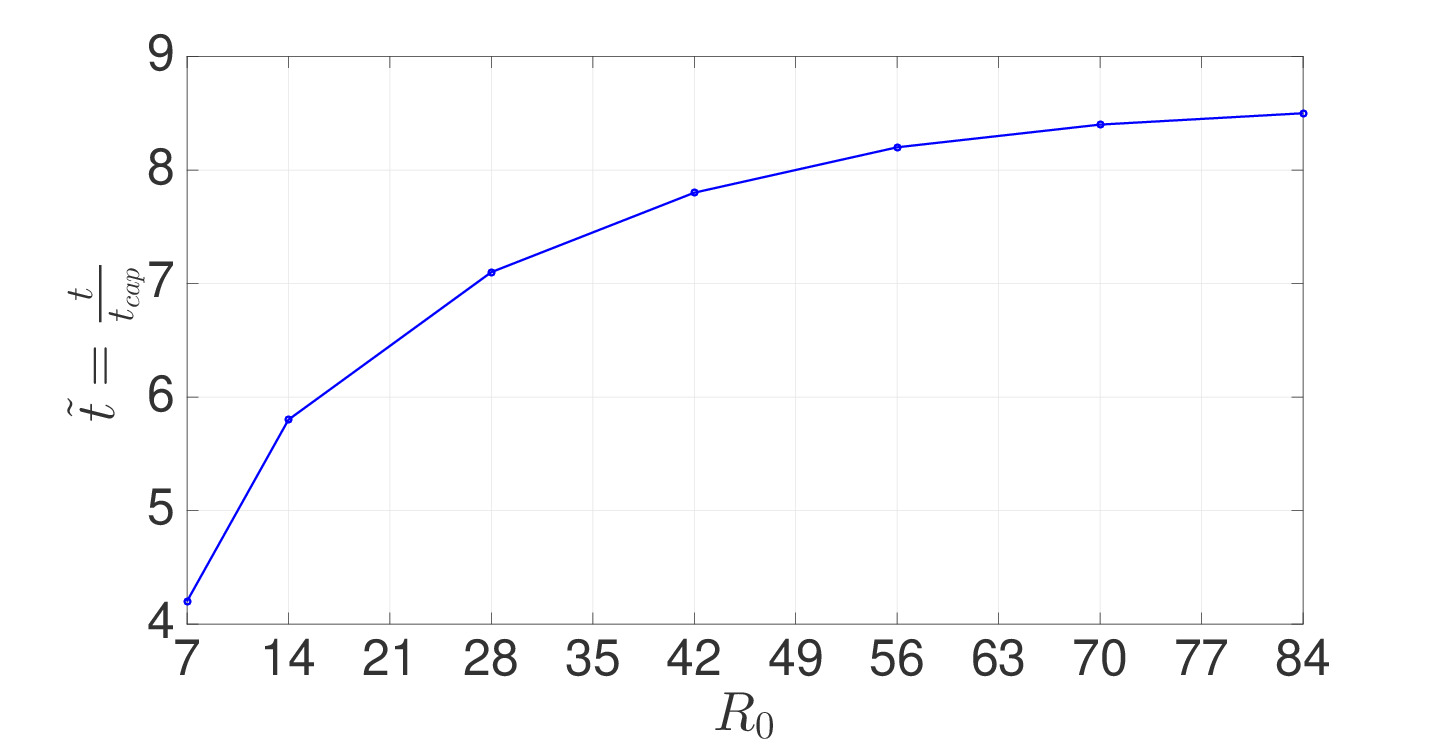}
\end{minipage}
\begin{minipage}{0.30\textwidth}
\centering
\includegraphics[width=0.93\columnwidth]{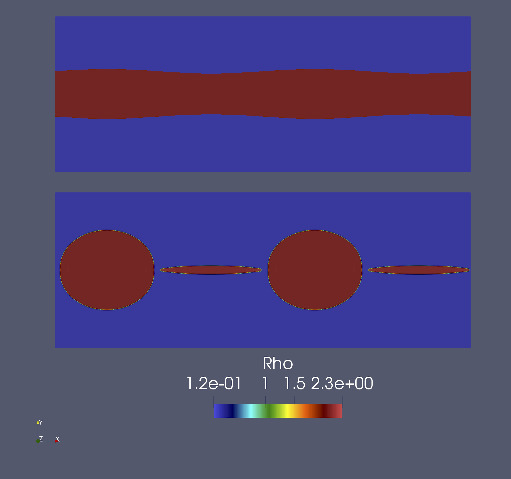}
\end{minipage}
\caption{Dimensionless breakup time as a function of ligament radius and two jet configurations (initial state and broken jet with satellite droplets formation).}
\label{fig:bu_time}
\end{figure}

In this section results for different ligament radii are presented. More specifically, the inter-particle strength $G_0$ has been fixed to 5.3, while the $Oh$ has been fixed to 0.1 and the ligament radius has been varied from 7 LU to 84 LU. With these parameters, the following density ratios and surface tensions have been obtained for the pseudo-potential multiphase model.
\begin{table}[h]
\caption{Densities (liquid and vapour) and surface tension as a function of ligament radius.}
\label{tab:radius}
\tabcolsep7pt\begin{tabular}{l c c c c c c c}
\hline
 $R_0$ & 7 & 14 & 28 & 42 & 56 & 70 & 84 \\
 \hline
$\rho_l$ & 2.112 & 2.117 & 2.125 & 2.135 & 2.145 & 2.156 & 2.167\\
$\rho_v$ & 0.081 & 0.084 & 0.090 & 0.097 & 0.106 & 0.116 & 0.126\\
$\sigma$ & 0.05159 & 0.05345 & 0.05775 & 0.06277 & 0.06848 & 0.07491 & 0.08206\\
\hline
\end{tabular}
\end{table}
With the parameters reported in Table \ref{tab:radius} the following dimensionless breakup time (normalized by means of the capillary time) may be obtained, and reported in Fig. \ref{fig:bu_time}.
As can be observed from Fig. \ref{fig:bu_time}, the breakup time is showing an asymptotic behaviour with increasing the ligament radius. This may lead to strong limitations in terms of grid-size requirement when hydrodynamic time-dependent phenomena has to be analysed with the proposed method. Future work is the comparison of presented results with outcomes of other collision operators, e.g. the Multi Relaxation Time (MRT), which would allow to keep the density ratio constant with varying the kinematic viscosity. Moreover, both SRT and MRT asymptotic results should be compared with predictions coming from sharp interface hydrodynamics.
\section{ACKNOWLEDGEMENTS}
The numerical simulations were performed on \textit{Zeus} HPC facility, at the University of Naples ``Parthenope''; \textit{Zeus} HPC has been realized through the Italian Government Grant PAC01$\_$00119 ``MITO - Informazioni Multimediali per Oggetti Territoriali'', with Prof. E. Jannelli as the Scientific Responsible. \\
This project has received funding from the European Union’s Horizon 2020 research and innovation programme under the Marie Sklodowska-Curie grant agreement No 642069.

\nocite{*}
\bibliographystyle{aipnum-cp}
\bibliography{sample}

\end{document}